\documentclass[twocolumn]{autart}    

\usepackage{graphicx}          
\usepackage{epstopdf}
\usepackage{amsmath}
\usepackage{amsfonts}
\usepackage{amssymb}
\usepackage{multicol}
\usepackage{stfloats}
\usepackage{verbatim}
\usepackage{mathrsfs}
\usepackage[longnamesfirst]{natbib}
\usepackage[colorlinks,linkcolor=red,anchorcolor=blue,citecolor=blue]{hyperref}

\theoremstyle{definition}
\newtheorem{theorem}{\textbf{Theorem}}

\newtheorem{assumption}{\textbf{Assumption}}

\newtheorem{lemma}{\textbf{Lemma}}

\newtheorem{remark}{\textbf{Remark}}

\newtheorem{corollary}{\textbf{Corollary}}
\newtheorem{definition}{\textbf{Definition}}

\newcommand*{\QEDA}{\hfill\ensuremath{\blacksquare}}

\begin{document}

\begin{frontmatter}

\title{ {Distributed stabilization control of rigid formations   with prescribed orientation}} 

\thanks[footnoteinfo]{The material in this paper   was  partially  presented at the IEEE Multi-Conference on Systems and Control, September 21-23, 2015, Sydney, Australia (\cite{sun2015orientation}) and the 54th IEEE Conference on Decision and Control, December 15-18, 2015, Osaka, Japan (\cite{myoung2015CDC}). }

\author[ANU]{Zhiyong Sun}\ead{zhiyong.sun@anu.edu.au},    
\author[GIST]{Myoung-Chul Park}\ead{mcpark@gist.ac.kr},               
\author[ANU]{Brian D. O. Anderson}\ead{brian.anderson@anu.edu.au},   
\author[GIST]{Hyo-Sung Ahn}\ead{hyosung@gist.ac.kr}

\address[ANU]{ National ICT Australia and
Research School of Engineering,  Australian National University, Canberra ACT
0200, Australia}  
\address[GIST]{School of Mechatronics, Gwangju Institute of Science and Technology, Gwangju, Republic of Korea}             

\begin{keyword}                           
Formation control; formation orientation; coordinate system,  rigidity theory.               
\end{keyword}                             

\begin{abstract}                          
Most rigid formation  controllers reported in the literature aim to only stabilize a rigid formation shape, while the formation orientation is not controlled.  This paper studies the problem of controlling rigid formations with   prescribed orientations   in both  2-D and 3-D spaces.   The proposed controllers involve  the commonly-used  gradient descent control for shape stabilization, and an additional term to control the directions of certain relative position vectors associated with certain chosen agents.  In this control framework, we show the minimal number of agents which should have  knowledge of a global coordinate system (2 agents for a 2-D rigid formation and 3 agents  for a 3-D rigid formation), while   all other agents do not require any global coordinate knowledge or any coordinate frame alignment to implement the proposed control.
The exponential convergence to the desired rigid shape and formation orientation is also proved.   Typical simulation examples are shown to support the analysis and performance of the proposed formation controllers.
\end{abstract}

\end{frontmatter}

\section{Introduction}
\subsection{Background and motivation}
Formation control for a group of autonomous mobile
agents has gained much attention due to its broad applications
in many areas including both civil and military fields.
A key problem in this domain that receives particular interest
is how to stabilize and maintain a geometrical formation shape  in
a distributed manner.
In the recent survey paper \cite{oh2014survey}, different types of formation control strategies are reviewed and compared, among which two most commonly-used approaches are
\begin{itemize}
\item the linear displacement-based approach: the desired formation is specified by a certain set of inter-agent \emph{relative positions} which means that the orientation of the final formation is implicitly fixed;
\item the nonlinear distance-based approach: the desired formation is specified by a certain set of inter-agent \emph{distances}, and the orientation of the target formation  is \emph{not} implicitly or explicitly defined.
\end{itemize}
For the first approach, all the agents must have their   coordinate bases with the same orientation (while the origins may be different) such that the desired relative position vectors are well defined and controlled between agents (see e.g. \cite{ren2008distributed, xiao2009finite}).  This means that \emph{all} the agents should be equipped with  compass to guarantee their coordinate orientation alignments, which may not be practical in e.g. compass-denied environment. The coordinate frame requirement was largely ignored in early works on formation control (as reviewed in \cite{oh2014survey}). It is only in recent years that the importance of coordinate frame issue has been recognized  in formation controller design and implementation.
In the case that   initially all the agents in the 2-D plane have different  local coordinate frames, one needs to design a combined control establishing  coordinate frame direction alignment and linear displacement-based formation stabilization   to ensure the convergence of a target shape \cite{Oh2014Orientation}. Furthermore,  it has also been shown in \cite{Meng2014Compass} that the assumption that all the agents have coordinate systems with the same orientation may not be realistic in practice as small perturbations in their local coordinate systems will cause unexpected behaviors for the displacement-based formation system. Thus in practice,   a coordinate-free formation control system is always favorable.  In \cite{aranda2015coordinate}, a coordinate-free formation control strategy was proposed by including a rotation matrix in the formation controller. The advantage of the coordinate-free property of the proposed formation controller in \cite{aranda2015coordinate} is paid by the price that the relative position measurements from \emph{all} other agents should be available to each individual agent, which implies that the coordinate-free formation control in \cite{aranda2015coordinate} is not  a distributed one.   Recent efforts also show that the bearing-based approach is another promising strategy to achieve a desired formation \cite{zhao2014bearing}. We note that  such an approach however still does not resolve the strict requirement of the global knowledge of coordinate frame orientation for individual agent.

All these disadvantages on the coordinate frame requirement can be avoided in the distance-based formation setup. This is because that in the  distance-based setup any global  coordinate system defining  a common orientation for all individual agents' coordinate frames is not required, and each agent can use its local coordinate basis to achieve a rigid formation shape {\color{blue} (we refer the readers to Fig. 3 in \cite{oh2014survey} for a comparison of coordinate basis requirement for these two approaches). }
  Rigid formation control has been discussed extensively in the literature, most of which has focused on the convergence analysis of formation shapes (see e.g. \cite{krick2009stabilisation}, \cite{anderson2014counting}, \cite{cortes2009global}, \cite{dorfler2010geometric}, \cite{oh2011formation}, \cite{tian2013global}, \cite{cai2015adaptive}).   Note that in many applications involving multi-agent coordination, a formation with both a desired shape and a particular orientation is required.
However, for distance-based rigid formation  control, the orientation   of the final formation is not controlled and actually not well defined, \footnote{We  need to distinguish different meanings of \emph{orientation} in the context of formation control. By regarding a rigid formation as a rigid body, the formation orientation relates to the overall rigid formation.
The orientation concept in e.g. \cite{Oh2014Orientation, Montijano2014} refers to the  orientation of the \emph{local coordinate frame} for each agent. We will distinguish different meanings by referring explicitly to either formation orientation or coordinate orientation.  Another orientation concept refers to the definition of signed area for a closed curve formed by a formation shape with a specific ordering of all agents (e.g. a triangle with positive/negative area). This concept will not be used in this paper. } which may limit the practical application of  shape controllers discussed in these previous works. In this paper, we aim to design distributed formation controllers to  achieve a desired rigid formation with a prescribed formation orientation.


\subsection{Related work}
The stabilization control of rigid formations with desired orientation was discussed in  \cite{pais2009formation} by using the  tensegrity theory and a projected collinear structure. However, the approach still requires all the agents to have knowledge of the orientation of a common   reference frame.
The problem of stabilizing only the orientation of rigid objects  subject to distance constraints  was studied in \cite{wang2011distributed}, \cite{markdahl2012distributed}, by assuming that the rigid shapes remain constant which  are not stabilized. Thus, the approaches in \cite{wang2011distributed} and \cite{markdahl2012distributed}  cannot be applied to solve the formation stabilization control task in question. In our previous paper \cite{Sun2014CDC} we showed a feasible approach to move or re-orient a rigid formation to a desired orientation by introducing distance mismatches; however, such orientation control approach, which is a by-product of the mismatched formation control problem,  indicates that the final formation is slightly distorted compared to the desired formation. Furthermore, the orientation control in \cite{Sun2014CDC} also requires global information in terms of all other agents' positions which is contrary to the formation control task  using a distributed approach.

In this paper we   propose feasible and distributed controllers to achieve both rigid shape stabilization and formation orientation control with \emph{minimal knowledge of global coordinate orientation} for the agent group. The basic idea underlying the controller design is to choose certain agents as \emph{orientation agents} (definitions will become clear in the context), for which some of the associated relative position vectors should achieve both  desired distances and directions specified in the global coordinate frame. We note that a very  general control framework for stabilizing an \emph{affine} formation was recently proposed in \cite{Lin_affine_formation}, in which a strict assumption  that the target formation should be \emph{globally rigid} was imposed to generate a rigid shape with orientation constraint. Such an assumption is not required in the control strategy proposed in this paper. {\color{blue} Also note that the formation orientation problem discussed in this paper is a stabilization control problem (i.e. to achieve a \emph{static} formation with desired orientation), while a motion generation problem involving rigid formation orientation was discussed in \cite{Hector2016maneuvering} with a totally different control architecture. }

Some preliminary results   were presented  in \cite{myoung2015CDC} and \cite{sun2015orientation}. This paper extends the results reported in   \cite{myoung2015CDC} and \cite{sun2015orientation}, by    providing a general and systematic approach  to solve this control problem with a minimal number of orientation agents. Compared to \cite{myoung2015CDC} and \cite{sun2015orientation}, the main extensions and contributions in this paper can be summarized as follows. First, the  results to be discussed in this paper can be applied to stabilize rigid shapes and orientations without any restriction on  agent numbers and  ambient space dimensions, while \cite{myoung2015CDC} presented preliminary results focusing   on proving asymptotic stability for a
2-D four-agent formation system.  Second,  we have removed the assumption that the target formation shapes are minimally rigid, which was a key assumption in  \cite{myoung2015CDC} and \cite{sun2015orientation}.  Instead, by developing different approaches in the proofs, this paper only assumes that   target formation shapes are infinitesimally rigid. Furthermore, by exploring  several novel observations concerning  the rigidity matrix from   graph rigidity theory, we will also prove an \emph{exponential} convergence to the desired formation shape with specified orientations. {\color{blue} Note that the exponential stability renders the robustness property of the proposed formation control system in the presence small measurement errors or perturbations. }

\subsection{Paper structure and notations}

The remaining parts of this paper are organized as follows. In Section 2, we introduce some
background on graph and rigidity theory as well as the problem formulation. Certain novel results on graph rigidity theory will be shown in this section. Section 3 provides the main result.   Typical simulation results are shown in Section 4. Finally, Section 5 concludes
this paper. Proofs for some key lemmas are given in the Appendix.

\emph{Notations.} The notations used in this paper are fairly
standard. $\mathbb{R}^n$ denotes the $n$-dimensional Euclidean space. $\mathbb{R}^{m\times n}$
denotes the set of $m\times n$ real matrices. A matrix or vector transpose is denoted by a superscript $T$. The rank, image  and null space of a matrix $M$ are denoted by  $\text{rank}(M)$, $\text{Im}(M)$ and $\text{null}(M)$, respectively.
We use $\text{diag}\{x\}$ to denote a  diagonal matrix with the entries of a vector $x$ on its diagonal, and $\text{span}\{v_1, v_2, \cdots, v_k\}$ to denote the subspace spanned by a set of vectors $v_1, v_2, \cdots, v_k$. The symbol $I_n$ denotes the $n \times n$ identity matrix. Let  $\mathbf{1}_n$ and  $\mathbf{0}_n$ denote  an $n$-tuple column vector of all ones and all zeros, respectively. When the subscripts are omitted, their dimensions should be clear in the context.   The notations $\otimes$ and $\wedge$  represent  the Kronecker product and cross product, respectively.

\section{Preliminaries and problem setup}
\subsection{Preliminary on graph theory}
Since formations of $n$ mobile agents are best described in terms of graph theory, we give a brief description of some of the
basic  definitions and facts needed.
Consider an undirected  graph with $m$ edges and $n$ vertices, denoted by $\mathcal{G} =( \mathcal{V}, \mathcal{E})$  with vertex set $\mathcal{V} = \{1,2,\cdots, n\}$ and edge set $\mathcal{E} \subset \mathcal{V} \times \mathcal{V}$.  The neighbor set $\mathcal{N}_i$ of node $i$ is defined as $\mathcal{N}_i: = \{j \in \mathcal{V}: (i,j) \in \mathcal{E}\}$.
  The matrix relating the nodes to the edges is called the incidence matrix $H = \{h_{ki}\} \in \mathbb{R}^{m \times n}$, whose entries are defined as (with arbitrary edge orientations for   \emph{undirected} formations considered here)
     \begin{equation}
     h_{ki} =  \left\{
       \begin{array}{cc}
       1,  &\text{ the } k\text{-th edge sinks at node }i  \\ \nonumber
       -1,  &\text{ the } k\text{-th edge leaves  node }i  \\ \nonumber
       0,  & \text{otherwise}  \\
       \end{array}
      \right.
      \end{equation}
For a connected and undirected graph, one has $\text{rank}(H) = n-1$ and $\text{null}(H) = \text{span}\{\mathbf{1}_n\}$.

\subsection{Rigidity theory}
Given a vertex element $i\in {\mathcal V}$ we associate to it a point
$p_i$ of Euclidean space $\mathbb{R}^d$. \footnote{In this paper we will focus on 2-D and 3-D formations, i.e.  $d = 2,3$.} The column vector $p=[p_1^T,p_2^T,\ldots,p_n^T]^T$
thus describes a \textit{framework} $(\mathcal{G}, p)$ of $n$ agents, labelled by the set
of vertices of $\mathcal {G}$. For any edge $k\in {\mathcal E}$ with head
$j$ and tail $i$ which is consistent with the construction of the matrix $H$, consider the associated relative position vector defined as $z_k=p_j-p_i$. Let
\begin{align*}
z& = [z_1^T,z_2^T,\cdots,z_m^T]^T \in \mathbb{R}^{dm} \\
Z(z)&=\text{diag} (z_1,z_2,\cdots,z_m) \in \mathbb{R}^{dm\times m}
\end{align*}
denote the associated column vector and block diagonal matrix, respectively. Note that there holds
\begin{align} \label{eq:definition_of_z}
z = (H \otimes I_d)p
\end{align}

With this notation at hand, we consider the
smooth distance map
\begin{equation}\label{eq:rigidity1}
r_{\mathcal{G}}:\mathbb{R}^{dn}\longrightarrow \mathbb{R}^{m},
r_{\mathcal{G}}(p)=(\|p_i-p_j \|^2)_{(i,j)\in {\mathcal E}}=Z(z)^Tz.
\end{equation}
The rigidity of frameworks is then defined as follows. \\
\begin{definition}  (\cite{asimow1979rigidity}) A framework $(\mathcal{G}, p)$ is rigid in $\mathbb{R}^d$ if there exists a neighborhood $\mathbb{U}$ of $p$ such that $r_{\mathcal{G}}^{-1}(r_{\mathcal{G}}(p))\cap \mathbb{U} = r_{\mathcal{K}}^{-1}(r_{\mathcal{K}}(p))\cap \mathbb{U}$ where $\mathcal{K}$ is the complete graph with the same vertex set as $\mathcal{G}$.
\end{definition}

Two frameworks $(\mathcal{G}, p)$  and $(\mathcal{G}, \bar p)$ are equivalent if $r_{\mathcal{G}}(p) = r_{\mathcal{G}}(\bar p)$ and are congruent if $\|p_i  - p_j\| = \|\bar p_i - \bar p_j\|$ for all $i,j \in \mathcal{V}$. A useful tool to study graph rigidity is the
{\bf rigidity matrix}, which  is defined as the Jacobian matrix
$R(p)=\frac{1}{2}\partial{{r_{\mathcal{G}}}(p)}/\partial (p)$. By inspection, $R(p)$  is an $m\times dn$
matrix given as
\begin{equation} \label{eq:rigidity_matrix}
R(p)=Z(z)^T(H \otimes I_d)
\end{equation}
Note that the entries of $R(p)$ only involve relative position vectors $z$, and we can rewrite it as $R(z)$.
The rigidity matrix will be used to determine the infinitesimal rigidity of a framework, as shown in the following theorem. \\

\begin{theorem} \label{theorem:rigidity}
 (\cite{hendrickson1992conditions}) Consider a framework   $(\mathcal{G}, p)$ in $d$-dimensional space with $n \geq d$ vertices and $m$ edges. It is infinitesimally rigid  if and only if
\begin{equation}
rank(R(p)) = dn-d(d+1)/2
\end{equation}
\end{theorem}
Specifically, the framework   $(\mathcal{G}, p)$ is infinitesimally rigid in $\mathbb{R}^2$ (resp. $\mathbb{R}^3$) if and only if $\text{rank}(R(p)) = 2n-3$ (resp.  $\text{rank}(R(p)) = 3n-6$). Obviously, in order to have an infinitesimally rigid framework, the graph should have at least $2n-3$ (resp. $3n-6$) edges in  $\mathbb{R}^2$ (resp. $\mathbb{R}^3$).

%
From Theorem \ref{theorem:rigidity}, one knows that the dimension of the null space of $R(p)$ for an infinitesimally rigid framework $(\mathcal{G}, p)$ in the $d$-dimensional space is   $d(d+1)/2$. The following Lemma characterizes the structure of its null space.
\begin{lemma} \label{lemma:null_vector}
(Null space of the rigidity matrix)
Suppose the framework $(\mathcal{G}, p)$ is infinitesimally rigid with  the associated rigidity matrix   denoted as $R(p)$.
\begin{itemize}
\item The $d = 2$ case: The null space  of $R(p)$  is of dimension 3 and is described as $\text{null}(R(p)) = \text{span}(q_1, q_2, q_3)$, where
\begin{align} \label{eq:basis_nullspace_2}
q_1 &= {\bf{1}}_n \otimes
\left[
\begin{array}{c}
1 \\
0\\
\end{array}
\right];
q_2 = {\bf{1}}_n \otimes
\left[
\begin{array}{c}
0 \\
1\\
\end{array}
\right];  \nonumber \\
q_3 &= [(K_3p_1)^T, (K_3p_2)^T, \cdots, (K_3p_n)^T]^T, \nonumber
\end{align}
and the matrix $K_3$ is defined as
\begin{align}  \nonumber
K_3 =
\left[
\begin{array}{cc}
0 & 1 \\
-1 & 0\\
\end{array}
\right]
\end{align}

\item The $d = 3$ case: The null space  of $R(p)$  is of dimension 6 and is described as $\text{null}(R(p)) = \text{span}(q_1, q_2, q_3, q_4, q_5, q_6)$, where
\begin{align} 
q_1 &= {\bf{1}}_n \otimes
\left[
\begin{array}{c}
1 \\
0\\
0\\
\end{array}
\right];
q_2 = {\bf{1}}_n \otimes
\left[
\begin{array}{c}
0 \\
1\\
0\\
\end{array}
\right];
q_3 = {\bf{1}}_n \otimes
\left[
\begin{array}{c}
0 \\
0\\
1\\
\end{array}
\right];  \nonumber \\
q_i &= [(K_ip_1)^T, (K_ip_2)^T, \cdots, (K_ip_n)^T]^T, i = 4,5,6;  \nonumber
\end{align}
and the matrix $K_i$ is defined as
\begin{align}  \nonumber
K_4 =
\left[
\begin{array}{ccc}
0 & 0 & 0 \\
0 & 0 & -1 \\
0 & 1 & 0 \\
\end{array}
\right];
K_5 =
\left[
\begin{array}{ccc}
0 & 0 & 1 \\
0 & 0 & 0 \\
-1 & 0 & 0 \\
\end{array}
\right];
K_6 =
\left[
\begin{array}{ccc}
0 & -1 & 0 \\
1 & 0 & 0 \\
0 & 0 & 0 \\
\end{array}
\right]
\end{align}
\end{itemize}
\end{lemma}

The proof and detailed analysis on how to  construct  the above null vectors can be found in the appendix. {\color{blue} In rigidity theory, any motion that lives in the null space of the rigidity matrix for an infinitesimally rigid framework is called an infinitesimal motion consisting of Euclidian motion (see e.g. \cite{tay1985generating}, \cite{anderson2010formal}). The structure of the null space of a rigidity matrix was shown   in e.g. \cite[Theorem   2.16]{Zelazo01012015}. The reason that we provide hereby an alternative proof  is to show a unified and clearer structure of its null space and the corresponding translational motion and rotational motion. Such a clear structure on the null space analysis will be helpful and useful in the controller design and stability analysis, as will be shown in the main part of this paper. }

The infinitesimal rigidity also guarantees the following property for a framework.
\begin{lemma} \label{lemma:noncollinear_noncomplanar}
Suppose the framework $(\mathcal{G}, p)$ is infinitesimally rigid. Then for any node $i$, the set of relative position vectors $p_j - p_i$, $j \in \mathcal{N}_i$ cannot all be collinear (in the 2-D case) or  all be coplanar (in the 3-D case).
\end{lemma}

The proof  can be found in the appendix. Lemma \ref{lemma:noncollinear_noncomplanar} will be useful for defining a feasible target formation by choosing some adjacent edges associated with certain agents (which will be discussed in Section 3.1).
\subsection{Gradient-based formation controller and problem formulation}

 Let $d_{k_{ij}}$ denote the desired distance of edge $k$  which links agent $i$ and $j$.
  We further define
  \begin{equation} \label{eq:distance_error}
  e_{k_{ij}} = \|p_i - p_j\|^2 - (d_{k_{ij}})^2
   \end{equation}
   to denote the squared distance error for edge $k$. {\color{blue} For ease of notation  we may  use $e_k$ and $d_k$ interchangeably in the sequel. This will also apply to  $d_{k_{ij}}$ and $d_{k}$, $z_{k}$ and   $z_{k_{ij}}$  in the following context when the dropping out of the dummy subscript $ij$ in each vector  causes no confusion. }  if no confusion is expected. The squared distance  error vector is denoted by $e = [e_1, \, e_2, \, \cdots, e_m]^T$. In this paper, we suppose  that each agent is modeled by a single integrator $\dot p_i  = u_i$ where $u_i$ is the controller to be designed for achieving the formation control objective.

In \cite{krick2009stabilisation}, the following formation control system was proposed:
 \begin{equation} \label{eq:original_system}
\dot{p}_i=-\sum_{j\in \mathcal N_i} (\|p_i-p_j\|^2-d_{k_{ij}}^2)(p_i-p_j),
\,\, i=1,\ldots,n
\end{equation}
The above control describes  a steepest descent gradient flow of the following potential function
\begin{equation} \label{eq:POT}
V_1(p)=\frac{1}{4}\sum_{(i,j)\in \mathcal E} (\|p_i-p_j\|^2-d_{k_{ij}}^2)^2
\end{equation}
This potential function  \eqref{eq:POT} for rigid shape stabilization and the associated gradient flow \eqref{eq:original_system} have been extensively
studied in the literature (see e.g. \cite{krick2009stabilisation,  cortes2009global, dorfler2010geometric, cao2011maintaining, oh2014distance}, \cite{anderson2014counting}).
However,    the above control \label{eq:position_system} and its extensions studied in these previous papers only stabilize a rigid formation shape, while the orientation of the formation is not specified. In this paper we will consider the problem of how to simultaneously stabilize a rigid shape and achieve a  desired orientation  for a target formation.


\section{Main result}
\subsection{Target formation and control framework}
Before describing the controller design, we first discuss how to define a target formation with the given inter-agent distance and formation orientation constraints.
As mentioned in the above section, the commonly-used gradient-based controller \eqref{eq:original_system} does not control the orientation and there are certain degrees of freedom relating to rotations for a converged formation. Intuitively, by regarding the rigid formation as a rigid body and specifying certain directions of some chosen edges in a global coordinate frame,  the orientation of the overall rigid formation can be fixed.
This will be the basic idea in the definition of a target formation and the controller design discussed in the sequel.

For simplifying the controller design and implementation, we   choose one agent and  a certain  number of its neighboring agents   as the specified agents  to implement  the additional orientation control task, with the associated edges between them being assigned with both \emph{distance constraints} and \emph{orientation constraints}. We term these agents with the additional orientation control task as \emph{orientation agents}, and other agents as \emph{non-orientation agents}. Thus, the target formation is defined with inter-agent distance constraints for all the agents, and orientation constraint for the chosen   edges between orientation agents.

\begin{figure}
  \centering
  \includegraphics[width=75mm]{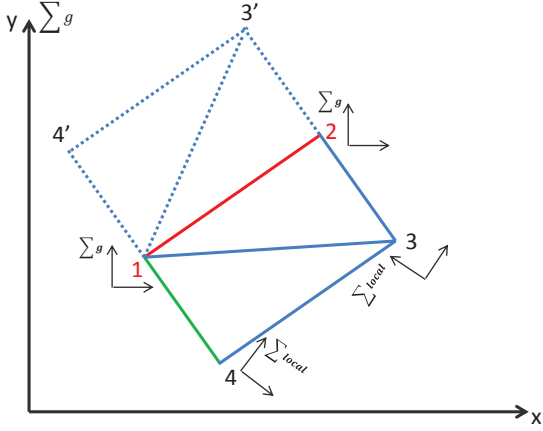}
  \caption{An example of controlling a 2-D rigid formation   with prescribed orientation.  Agent 1 and  one of its neighbors, agent 2, are chosen as orientation agents. The relative position vector $p_2-p_1$ associated with edge (1,2) is used to describe the desired orientation, which is denoted by red color (in this example $(1,2) \in \mathcal{E}_o$). }
  \label{Fig:orientation_2D}
\end{figure}

\begin{figure}
  \centering
  \includegraphics[width=75mm]{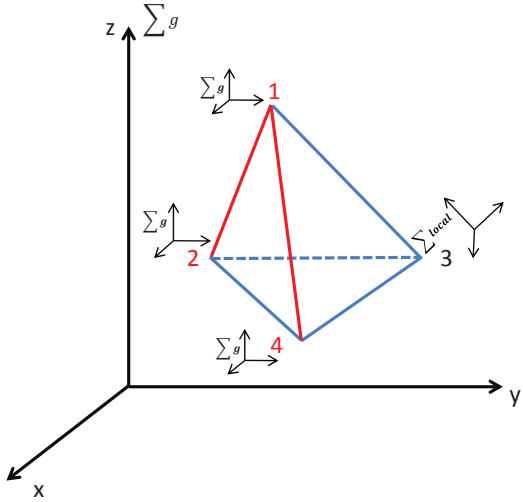}
    \caption{An example of controlling a 3-D rigid formation   with prescribed orientation.  Agent 1 and two of its neighbors, agents 2 and 4, are chosen as orientation agents. The relative position vectors $p_2-p_1$ and $p_4-p_1$ associated with edges (1,2) and (1,4) are used to describe the desired orientation, which are denoted by red color (in this example $(1,2), (1,4) \in \mathcal{E}_o$).  }
  \label{Fig:orientation_3D}
\end{figure}

For the convenience of later analysis, we   denote $\mathcal{G}_o$ as the underlying graph of the orientation control to distinguish it with the underlying graph  $\mathcal{G}$ of the formation shape control. If the edge $(i,j)$ associated with agent $i$ and $j$ is chosen in the orientation control in $\mathcal{G}_o$, we denote it as   $(i,j) \in \mathcal{E}_o$.
The  set of neighboring agents of orientation agent $i$ chosen in the orientation control  is defined as $\mathcal{N}_i^o: = \{j \in \mathcal{V}: (i,j) \in \mathcal{E}_o\}$. The desired direction for the relative position vector $p_j - p_i$ for edge $(i,j) \in \mathcal{E}_o$ is  denoted by a given vector $\hat p_{ji} : = \hat p_j - \hat p_i$.  Thus, the   orientation control is   to additionally stabilize the relative position $p_j - p_i$ to the desired one $\hat p_j - \hat p_i$ with $(i,j) \in \mathcal{E}_o$. Due to the rigid body property of a desired rigid formation, the formation orientation can be determined by the directions of a certain set of desired relative position vectors.
We show two examples,  a 2-D four-agent rectangular formation and a 3-D tetrahedral formation  depicted in Fig. \ref{Fig:orientation_2D} and Fig. \ref{Fig:orientation_3D}, respectively, to illustrate the formation control framework.

Note that   any two agents associated with one edge can be chosen as orientation agents, and there is no need to design a centralized algorithm   for the selection of  the orientation agents. To define a target formation with prescribed orientation, one can first choose one agent and then select one of its   non-collinear relative vectors  (for 2-D formations) or two of its non-coplanar relative vectors (for 3-D formations) to specify the desired formation orientation. According to Lemma \ref{lemma:noncollinear_noncomplanar}, such non-collinear or non-coplanar adjacent edges are guaranteed to exist for any agent to define a target formation. To sum up, we give a formal definition of a target formation.

\begin{definition} (Target formation)
The target formation is defined as $(\mathcal{G}, \tilde p)$ which satisfies the following constraints
\begin{itemize}
\item Distance constraints: $\|\tilde p_i - \tilde p_j\| = d_{k_{ij}}$, $\forall (i,j) \in \mathcal{E}$;
\item Orientation constraints: $\tilde p_i - \tilde p_j = \hat p_i - \hat p_j$, $\forall (i,j) \in \mathcal{E}_o$;
\end{itemize}
\end{definition}
Note that there should hold $\|(\hat p_j - \hat p_i)\| = d_{k_{ij}}$ so that the orientation constraint is consistent with the formation shape constraint.
In order to well define the orientation constraint,  we need the following assumption.
\begin{assumption}
All  orientation agents should be equipped with   coordinate systems with the same direction   aligned with the global coordinate system.
\end{assumption}

Take the formation control formulation in Fig. \ref{Fig:orientation_2D} as an example. Since agents 1 and 2 are chosen as   orientation agents, their coordinate systems should be aligned with the global coordinate system denoted by $\sum_g$. Such a global coordinate system is required to define the desired relative position vector $(\hat p_j  - \hat p_i)$ for $(i,j) \in \mathcal{E}_o$.  Thus Assumption 1 provides a necessary condition for the controller design and implementation.

\subsection{Discussions on  reflection ambiguity}

By specifying the direction of  one edge in a 2-D formation, there exists a reflected formation with the same prescribed orientation   (in the example in  Fig. \ref{Fig:orientation_2D}  the reflected formation, denoted by   dotted blue lines, is obtained by the reflection via the mirror  edge (1,2)). Such reflection ambiguity can be avoided by specifying the direction of an additional relative position vector (such as the one associated with edge (1,4)) or by assuming  that the initial formation shape
starts close to the desired one. In the latter case the formation shape will converge to the desired one
instead of converging to the reflected one (by the convergence property of the
\emph{gradient} property of the proposed formation control system, to be proved in Theorem \ref{theorem:gradient}). Similarly, in the 3-D case   there exists a reflected formation  via the   mirror plane spanned by that two chosen relative position  vectors (in the example in Fig. \ref{Fig:orientation_3D} the mirror plane is  spanned by that two relative vectors in edges (1,2), (1,4)). Such reflection ambiguity can be avoided  by specifying the direction of an additional relative position vector (such as the one associated with edge (1,3)), or by assuming  that the initial formation shape
starts close to the desired one. 

It might also be true that by setting the orientations for more edges may lead to a larger region of attraction, but this results in more orientation agents requiring the knowledge of the global coordinate system. Since in this paper we only focus on local convergence, the possibility of using more orientation edges will not be further exploited. In the current problem setting, the minimum number of orientation agents required in controller design is 2 (for 2-D rigid formations) and 3 (for 3-D rigid formations). Such minimum number will be formally proved in later analysis which also guarantees the local convergence to a target formation with desired shape and orientation.

\subsection{Controller design}
We propose the following formation stabilization controller:
\begin{align}\label{eq:rigid_system_orientation}
\dot p_i(t) = & \underbrace{ \sum_{j \in \mathcal{N}_i} (p_j(t) - p_i(t)) (\|p_j(t) - p_i(t)\|^2 - d_{k_{ij}}^2)}_{\text{shape control term, if} \,(i,j) \in \mathcal{E}} \nonumber \\
& + \underbrace{\sum_{j \in \mathcal{N}^o_i} \left((p_j(t) - p_i(t)) - (\hat p_j - \hat p_i) \right) }_{\text{orientation control term, if}\,  (i,j) \in \mathcal{E}_o}
\end{align}
It is obvious from Eq. \eqref{eq:rigid_system_orientation} that the proposed control is distributed since only local information from neighboring agents in terms of relative positions is needed. {\color{blue} In the later analysis we will also show that the overall system consisting of $n$ agents described by \eqref{eq:rigid_system_orientation} is a \emph{gradient} system associated with a cost function. }

The above formation control system \eqref{eq:rigid_system_orientation} can be written in a compact form
\begin{align} \label{eq:compact}
\dot p = -R^T e  - (L_o \otimes I_d) \bar p
\end{align}
where   $L_o$ is the Laplacian matrix of the underlying undirected graph $\mathcal{G}_o$ for the orientation control, and the vector $\bar p = [\bar p_1^T, \bar p_2^T, \cdots, \bar p_n^T]^T$ is  defined as
 $\bar p_i = p_i - \hat p_i$ if $i$ is a chosen orientation agent \footnote{ Note that the vector $\bar p_i$  is not an actual control input as $\hat p_i$ may not be available for agent $i$ (the actual control term is $\hat p_{ji} : = \hat p_j - \hat p_i$). The introduction of $\bar p_i$ is for the convenience of analysis and for writing a compact form of the formation system as in  \eqref{eq:compact}. },  or $\bar p_i = \bf{0}$ otherwise.

For the formation control system \eqref{eq:compact}, the set of the desired equilibrium is described as
\begin{align}  \label{eq:equilibrium_set}
\mathcal{M} = \{ p \in \mathbb{R}^{dn} |  e(p) =  \mathbf{0}, p_i-p_j = \hat p_i- \hat p_j, \forall (i,j) \in \mathcal{E}_o\}
\end{align}
which satisfies the constraints in Definition 2.

\textbf{Example}:
We show an example to illustrate the above controller design.
Suppose a group of four agents is tasked to achieve  a rigid   shape, with the additional orientation control assigned to edge $(1,2) \in \mathcal{E}_o$ , which is illustrated in Fig. \ref{Fig:orientation_2D}.  The formation control system takes the following form
\begin{align} \label{eq:example2D}
\dot p_1 = & e_{12}(p_2 - p_1) + e_{13}(p_3 - p_1) + e_{14}(p_4 - p_1) \nonumber \\
& + (p_2 - p_1) - (\hat p_2 - \hat p_1) \nonumber \\
\dot p_2 = & e_{12}(p_1 - p_2) + e_{23}(p_3 - p_2)  + (p_1 - p_2) - (\hat p_1 - \hat p_2) \nonumber \\
\dot p_3 = & e_{13}(p_1 - p_3) + e_{23}(p_2 - p_3) + e_{34}(p_4 - p_3)  \nonumber \\
\dot p_4 = & e_{14}(p_1 - p_4) + e_{34}(p_3 - p_4)
\end{align}
(note that in the above equations the subscript notation for $e$ is slightly different to previous sections, in that $e_{ij}$ here denotes the squared distance error associated with the edge $(i,j) \in \mathcal{E}$).
The Laplacian matrix $L_o$ for the underlying graph of orientation control is constructed as
\begin{align} L_o = \left[
\begin{array}{cccc}
1 & -1 & 0 & 0 \nonumber \\
-1 & 1 & 0 & 0 \nonumber \\
0 & 0 & 0 & 0 \nonumber \\
0 & 0 & 0 & 0 \nonumber \\
\end{array}
\right]
\end{align}
and the vector $\bar p$ is constructed as $\bar p = [(p_1 - \hat p_1)^T, (p_2 - \hat p_2^T, \mathbf{0}^T, \mathbf{0}^T]^T$. The formation system \eqref{eq:example2D} can then be written in the compact form shown in \eqref{eq:compact}. \QEDA

\subsection{Properties of the formation control system}


In the following, we show several properties of the proposed control \eqref{eq:rigid_system_orientation}.
\begin{lemma} \label{lemma:fixed_centroid}
The position of the formation centroid is preserved by the above control law \eqref{eq:rigid_system_orientation}.
\end{lemma}
\textbf{Proof}
Denote by $p_c \in \mathbb{R}^d$ the center of the mass of the formation, i.e.,
\begin{equation} \label{eq:p_c}
p_c = \frac{1}{n} \sum_{i=1}^n p_i = \frac{1}{n}(\mathbf{1}_n \otimes I_d)^T p
\end{equation}
One has
\begin{align}
\dot p_c = & \frac{1}{n}(\mathbf{1}_n \otimes I_d)^T \dot p  \nonumber \\
= & -\frac{1}{n}(\mathbf{1}_n \otimes I_d)^T (R^T e + (L_o \otimes I_d) \bar p) \nonumber \\
= & -\frac{1}{n}(\mathbf{1}_n \otimes I_d)^T (L_o \otimes I_d) \bar p  \nonumber \\
& -\frac{1}{n}\left( Z^T  (H \otimes I_d) (\mathbf{1}_n \otimes I_d) \right)^T e
\end{align}
Note that $(\mathbf{1}_n \otimes I_d)^T (L_o \otimes I_d) \bar p = ((\mathbf{1}_n^T L_o) \otimes I_d)\bar p = 0$ and $\left( Z^T  (H \otimes I_d) (\mathbf{1}_n \otimes I_d) \right)^T e = 0$ because $\text{null}(H) = \text{span}\{\mathbf{1}_n\}$. Thus $\dot p_c = 0$, which indicates that the position of the formation centroid remains constant.
\qed

\begin{lemma} \label{lemma:coordinate_I}
For all non-orientation agents, their local coordinate systems are sufficient to implement the control law.
\end{lemma}
\textbf{Proof}
Suppose agent $i$ is a non-orientation agent and its position in the global coordinate system $\sum_g$ is measured as $p_i^g$, while $p_i^i$, $p_j^i$ stand  for agent $i$ and its neighboring agent $j$'s positions, respectively, measured by agent $i$'s  local coordinate system. The controller for non-orientation agent $i$ can be written in its local coordinate as
 \begin{equation} \label{eq:local_coordinate_controller}
\dot p_i^i(t) =  \sum_{j \in \mathcal{N}_i} (p_j^i(t) - p_i^i(t)) (\|p_j^i(t) - p_i^i(t)\|^2 - d_{k_{ij}}^2)
 \end{equation}

 Clearly,  there exist a rotation matrix $\mathcal{Q}_i \in \mathbb{R}^{d \times d}$ and a translation vector $\vartheta_i \in \mathbb{R}^{d}$, such that $p_j^i = \mathcal{Q}_i p_j^g + \vartheta_i$. We rewrite the  controller \eqref{eq:local_coordinate_controller} for the non-orientation agent $i$ in the global coordinate system $\sum_g$ as follows
\begin{align} \nonumber
\dot p_i^g &= u_i^g =\mathcal{Q}_i^{-1}u_i \\ \nonumber
&=   \mathcal{Q}_i^{-1} \sum_{j \in \mathcal{N}_i}  e_{k_{ij}}^i  ( p_j^i -  p_i^i)  \\ \nonumber
&= \sum_{j \in \mathcal{N}_i} e_{k_{ij}}^g \mathcal{Q}_i^{-1} \mathcal{Q}_i ( p_j^g -  p_i^g) \\
&= \sum_{j \in \mathcal{N}_i} e_{k_{ij}}^g  ( p_j^g -  p_i^g) \label{original_controller}
\end{align}
which has the same form as \eqref{eq:local_coordinate_controller}.  Since $\mathcal{Q}_i$ and $\vartheta_i$  are chosen arbitrarily, the above equation indicates that the designed controllers for non-orientation agents are independent of the global coordinate basis.
\qed

To avoid notation complexity we omit the superscript in other parts of this paper for the convenience of
analysis.
This controller property has been illustrated in Fig. \ref{Fig:orientation_2D}   and Fig. \ref{Fig:orientation_3D}.  In the example shown in Fig. \ref{Fig:orientation_2D}, agent  3 or 4 is not an orientation agent and its   coordinate system orientation  does not need to be aligned with the global coordinate system. As a consequence of Lemma \ref{lemma:coordinate_I}, the minimum number of orientation agents is 2   for a  2-D rigid formation and 3  for a  3-D rigid formation, which guarantees a minimum knowledge of global coordinate frame for the multi-agent formation group.

\subsection{Convergence analysis}
We will first show that the gradient property of the proposed controller and a general result on the convergence.

\begin{theorem} \label{theorem:gradient}
 The formation control system with the proposed controller \eqref{eq:rigid_system_orientation} describes a gradient control system and the formation system converges to the  largest invariant set in the set $\mathcal{O}(z)$ defined as
\begin{align} \label{eq:larget_invariant_set}
\mathcal{O}(z)  = \{z | R(z)^T e(z)+ (L_o \otimes I_d) \bar p(z) = 0 \}
\end{align}
\end{theorem}

\textbf{Proof}
We choose the same potential function in \eqref{eq:POT} as the potential for the shape control,
and the following potential function
\begin{align}
V_2 & = \frac{1}{2} \sum_{(i,j) \in \mathcal{E}_o}  \| (p_j(t) - p_i(t)) -  (\hat p_j  - \hat p_i)  \|^2  \nonumber \\
& = \frac{1}{2}\bar p^T (L_o \otimes I_d) \bar p
\end{align}
for the orientation control.
The composite potential function is then defined as   $V = V_1 +V_2$.


The dynamical system for the relative position $z$ defined in \eqref{eq:definition_of_z} is
\begin{align} \label{eq:compact_z}
\dot z & = (H \otimes I_d) \dot p  \nonumber \\
& = -(H \otimes I_d) R^T e  - (H \otimes I_d) (L_o \otimes I_d) \bar p
\end{align}

and the distance error system is described by
\begin{align} \label{eq:compact_e}
\dot e = 2R \dot p  = -2RR^T e  - 2R (L_o \otimes I_d) \bar p
\end{align}
Note that the potential functions $V_1$ and $V_2$ are functions involving only relative position vectors in terms of $z$ and $e$ rather than the absolute position vector $p$. \footnote{ Also note that the distance error vector $e$ can be written in terms of $z$ according to the definition of $e$ in \eqref{eq:distance_error}. } Thus, we can write the potential as $V(z)$ for the self-contained $z$ system \eqref{eq:compact_z}.
We then calculate the derivative of the potential $V_1$ and $V_2$ along the trajectories of system \eqref{eq:compact_z} and \eqref{eq:compact_e}:
\begin{align}
\dot V_1 &= \frac{1}{2}e^T \dot e  = e^T (-RR^T e  - R (L_o \otimes I_d) \bar p) \nonumber \\
&  = -e^T RR^T e  -  e^T R (L_o \otimes I_d) \bar p
\end{align}
and
\begin{align}
\dot V_2 &= \bar p^T  (L_o \otimes I_d) \dot {\bar p} \nonumber \\
&= \bar p^T (L_o \otimes I_d) (-R^T e  - (L_o \otimes I_d) \bar p) \nonumber \\
&  = -\bar p^T (L_o \otimes I_d) R^T e  -  \bar p^T (L_o \otimes I_d)  (L_o \otimes I_d) \bar p
\end{align}
where in the second equality we have used the non-trivial result $(L_o \otimes I_d) \dot {\bar p} = (L_o \otimes I_d) \dot { p}$.
The derivative of $V$ can be computed as
\begin{align}
\dot V  = & \dot V_1  + \dot V_2 \nonumber \\
= & -e^T RR^T e  -  e^T R (L_o \otimes I_d) \bar p   \nonumber \\
& -\bar p^T (L_o \otimes I_d) R^T e  -  \bar p^T (L_o \otimes I_d) (L_o \otimes I_d) \bar p  \nonumber \\
= & -e^T RR^T e  -  2e^T R (L_o \otimes I_d) \bar p   \nonumber \\
& -  \bar p^T (L_o \otimes I_d)  (L_o \otimes I_d) \bar p  \nonumber \\
= & -\left (R^T e+ (L_o \otimes I_d) \bar p \right)^T  \left(R^T e+ (L_o \otimes I_d) \bar p \right)   \nonumber \\
\leq & 0
\end{align}
The above derivative calculation thus implies that the formation system \eqref{eq:rigid_system_orientation} describes a  gradient descent flow for the composite potential $V$. Furthermore, the sub-level set of the potential $V(z)$ is compact  with respect to the self-contained $z$ system \eqref{eq:compact_z}. By LaSalle Invariance Principle,
the solution of the formation system \eqref{eq:compact_z} converges to the largest invariant set in the set $\mathcal{O}(z)  = \{z| \dot V = 0\}$ described in \eqref{eq:larget_invariant_set}.  \qed



In general, a global picture of  convergence analysis  for a rigid formation control system is hard to obtain due to the existence of multiple equilibria (see discussions in e.g. \cite{anderson2014counting}). Because the proposed control is a gradient law, the set $\mathcal{O}(z)$ also describes the set of equilibrium points for \eqref{eq:rigid_system_orientation}.  Note that the desired equilibria set $\mathcal{M}$ is a subset of   $\mathcal{O}(z)$.
Similar to most works on rigid formation stabilization, in the following    we will   focus on local convergence analysis. In particular, we aim to show that the convergence to the target formation with desired distances and orientation is \emph{exponentially fast}.
The   analysis is based on the linearization technique. We first compute the Jacobian of the vector field in the right-hand side of \eqref{eq:compact} around a desired equilibrium $\tilde p \in \mathcal{M}$:
\begin{align}
J_f & = \frac{\partial (-R^T e  - (L_o \otimes I_d) \bar p)}{\partial p} |_{p = \tilde p }  \nonumber \\
& = - \frac{\partial R^T}{\partial p} e |_{p = \tilde p } - R^T \frac{\partial e}{\partial p} |_{p = \tilde p } - \frac{ \partial (L_o \otimes I_d)     \bar  p  }{\partial p} |_{p = \tilde p } \nonumber \\
& = - (R^T R + (L_o \otimes I_d))|_{p = \tilde p }
\end{align}
where $\frac{\partial R^T}{\partial p} e(p) |_{p = \tilde p } = \mathbf{0}$ due to the fact that $e(p) = \mathbf{0}$ for a point $p$ in the equilibrium set $\mathcal{M}$, and $\frac{\partial e}{\partial p} |_{p = \tilde p } = R$ according to the definition of the rigidity matrix.

Thus, the linearization equation of  \eqref{eq:compact}  is described as
\begin{align} \label{eq:linearization_p}
\delta \dot p =  - (R^TR  +  L_o \otimes I_d) \delta p
\end{align}

In the following, we prove that the convergence is exponentially fast.
\begin{theorem} \label{theorem:exponential}
Suppose  the target formation is   infinitesimally rigid and initial positions of all the agents are chosen such that the initial formation is close to the desired formation.
With the proposed control law \eqref{eq:rigid_system_orientation}, the convergence to the correct formation shape and orientation is \emph{exponentially fast}.
\end{theorem}
Before giving the proof of the above result, we first show a key lemma on the property of the linearization matrix.
\begin{lemma} \label{lemma:null_vector_F}
Suppose the target formation is infinitesimally rigid and the orientation edges are selected according to  Section 3.1. Then the linearization matrix $\mathcal{F} := R^TR  +  L_o \otimes I_d$ is positive semidefinite and has $d$ zero eigenvalues. Furthermore, there holds $\text{null}(\mathcal{F}) = \text{null}(H \otimes I_d) = \text{span} ({\bf {1}}_n \otimes I_d )$.
\end{lemma}

The proof can be found in the appendix.

\textbf{Proof of Theorem \ref{theorem:exponential}}\,\,\,\,
As shown  in Lemma \ref{lemma:fixed_centroid}, the formation centroid   is stationary. We construct an orthogonal matrix   $Q \in \mathbb{R}^{dn \times dn}$ whose   first $d$ rows   are $\frac{1}{\sqrt{n}}(\mathbf{1}_n \otimes I_d)^T$. With $Q$, one can perform the coordinate transform on $p$ as
\begin{align} \label{eq:coordinate_transform}
\tilde p = Q p = \left[
\begin{array}{c}
p^o \\
p_r
\end{array}
\right]
\end{align}
where $p^o  = \sqrt{n} p_c$ according to the definition of $p_c$ in \eqref{eq:p_c} and the structure of $Q$. From Lemma \ref{lemma:fixed_centroid}, one has $ \dot p^o = \sqrt{n} \dot p_c = \bf 0$. We also define a reduced transformation matrix $Q_r \in \mathbb{R}^{d(n-1) \times dn}$, obtained from $Q$ by removing the first $d$ rows. Note that there holds $p = Q^{-1} \tilde p = Q^{T} \tilde p$ and $p_r = Q_r p$. For the linearized system \eqref{eq:linearization_p}, one can obtain the following coordinate-transformed  system
\begin{align} \label{eq:coordinate_trans}
\left[
\begin{array}{c}
\delta \dot p^o \\
\delta \dot p_r
\end{array}
\right] = Q \dot \delta p & =  - Q (R^TR  +  L_o \otimes I_d) \delta p \nonumber \\
& :=  - Q \mathcal{F} Q^{-1} \delta \tilde  p
\end{align}
According to the structure of the matrix $Q$, there holds
\begin{align}
Q \mathcal{F} Q^{-1} &=
\left[
\begin{array}{c}
\frac{1}{\sqrt{n}}(\mathbf{1}_n \otimes I_d)^T \\
Q_r
\end{array}
\right]
\mathcal{F} \left[\frac{1}{\sqrt{n}}(\mathbf{1}_n \otimes I_d)\,\,\,\, Q_r^T \right] \nonumber \\
&=
\left[
\begin{array}{cc}
\bf 0 \,\,\,\,\,& \bf 0 \\
\bf 0 \,\,\,\,\,& Q_r \mathcal{F} Q_r^T
\end{array}
\right]
\end{align}
Therefore,

\begin{align} \label{eq:coordinate_trans2}
\delta \dot p^o &= \bf 0 \nonumber \\
\delta \dot p_r &= - Q_r \mathcal{F}  Q_r^T  \delta  p_r
\end{align}
According to  the definition of $Q_r$, the range space of $Q_r^T$ is the orthogonal complement of the subspace $\text{span} ({\bf {1}}_n \otimes I_d )$.  This, together with Lemma \ref{lemma:null_vector_F}, implies that the linearization matrix $-Q_r \mathcal{F}  Q_r^T$ is \emph{negative definite}. Thus the convergence to the origin for the system \eqref{eq:coordinate_trans2} is locally exponentially fast. Since the system \eqref{eq:coordinate_trans2} is obtained from the   system \eqref{eq:linearization_p} by a linear coordinate transformation  described  in \eqref{eq:coordinate_transform}, the above statement also implies that the convergence to a point in the desired equilibrium $\mathcal{M}$ for   the original system \eqref{eq:compact}  is \emph{locally exponentially fast} (\cite[Theorem 4.13]{khalil1996nonlinear}). For the linearized system, the guaranteed  exponential convergence rate obtained in the linearization analysis is $\gamma = \lambda_{\text{min}} (Q_r \mathcal{F}  Q_r^T)$. Note that by the Courant-Fischer Theorem \cite[Theorem 8.9]{zhang2011matrix}, the rate $\gamma = \lambda_{\text{min}} (Q_r \mathcal{F}  Q_r^T)$ is the same to the smallest positive eigenvalue of $\mathcal{F}$.
\qed

\begin{remark}
One may ask what happens if the formation is initially with a correct shape but needs to adjust the orientation by applying the designed controller \eqref{eq:rigid_system_orientation}. As can be seen from \eqref{eq:compact_e}, $e(0) = 0$ does not imply $\dot e(t) = 0$ when the proposed controller \eqref{eq:rigid_system_orientation} is applied. Thus, during the orientation adjustment the formation shape will be temporarily lost until the formation converges to the desired shape and orientation. If the formation shape should   remain unchanged during the orientation adjustment, the control action should live in the null space of the rigidity matrix with a target formation shape derived   in Lemma \ref{lemma:null_vector}. A sufficient condition for the controller design in this case is to ensure that the $z$ system   takes the form as $\dot z_i = \omega \times z_i$ (where $\omega$ indicates the angular velocity and $\times$ denotes the cross product) which guarantees a constant norm of $z$ and thus a preserved formation shape.
\end{remark}

{\color{blue}
\begin{remark}
In the above analysis we do not confine the formation to be minimally rigid (which is a commonly-used assumption in most literature on rigid formation control). Also we prove the local exponential convergence if the target formation is infinitesimally rigid (a more relaxed assumption than minimal rigidity). Exponential stability brings about several nice properties such as the robustness to system perturbations (e.g. measurement errors). This will be considered in future research, along the same research direction on robustness issues in rigid formation control \cite{SMA16TACsub}. Note that the exponential convergence cannot be directly extended to the general convergence to a set stated in Theorem \ref{theorem:gradient}. This is because, as indicated in the proof of Theorem \ref{theorem:exponential}, the local exponential convergence to a target formation depends on the maximum rank condition of the rigidity matrix of a target formation, and for other formations  defined in the set \eqref{eq:larget_invariant_set} one cannot guarantee that they are infinitesimally rigid.
\end{remark}

}

\section{Illustrative examples}
\begin{figure}
  \centering
  \includegraphics[width=75mm]{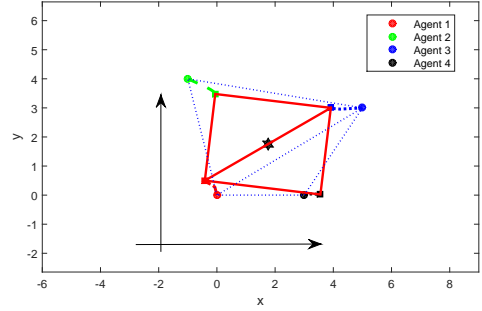}
  \caption{Stabilization of  a rigid rectangular formation without orientation control. The initial and final positions are denoted by circles and squares, respectively. The initial formation is denoted by dotted blue lines, and the final formation is denoted by red solid lines. }
  \label{fig:orientation_without}
\end{figure}

In this section we provide several simulations to show formation  behaviors and controller performance of the proposed control. Consider a 4-agent  formation system, with the desired distances given as $d_{12}^* =d_{34}^* = 3$, $d_{23}^* = d_{14}^* =4$, $d_{13}^* = 5$ corresponding to a rectangular shape. The initial positions for each agent are chosen as $p_1(0) = [0,\, 0]^T$, $p_2(0) = [-1,\, 4]^T$, $p_3(0) = [5,\, 3]^T$ and $p_4(0) = [3,\, 0]^T$, so that the initial formation shape is close to the target shape. When the conventional controller \eqref{eq:original_system}  is used,  the trajectories of each agent and the final shape are depicted in Fig. \ref{fig:orientation_without}, from which it can be seen that although the desired shape is achieved, the formation orientation is undefined.

\begin{figure}
  \centering
  \includegraphics[width=75mm]{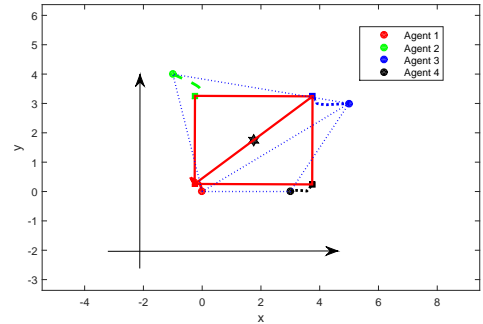}
  \caption{Stabilization of a rigid rectangular formation with prescribed orientation. The initial and final positions are denoted by circles and squares, respectively. The initial formation is denoted by dotted blue lines, and the final formation is denoted by red solid lines. The black star denotes the formation centroid. }
  \label{fig:orientation_with_centroid}
\end{figure}

\begin{figure}
  \centering
  \includegraphics[width=75mm]{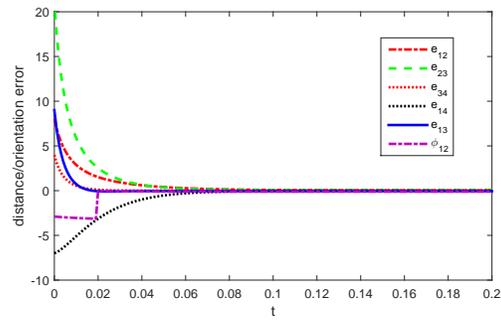}
  \caption{Convergence of the distance/orentation errors with the proposed controller \eqref{eq:rigid_system_orientation}. The orientation error $\phi_{12}$ is defined as $\phi_{12} = \text{arctan}((p_{2,y} - p_{1,y})/(p_{2,x} - p_{1,x})) - \pi/2$. }
  \label{fig:orientation_with_centroid_errors}
\end{figure}

We then consider the simulation using the proposed controller \eqref{eq:rigid_system_orientation}. We suppose the target formation should be the one with the rigid rectangular shape in addition that  the relative position vector $p_2 - p_1$ associated to edge $(1,2)$ should be aligned with  the direction of  the $y$-axis and the relative position vector $p_4 - p_1$ associated with the edge $(1,4)$ should be aligned with    the direction of  the $x$-axis in the global coordinate. The desired relative vector for edge $(1,2)$ is set as $\hat p_2 - \hat p_1 = (0,3)^T$ and the initial positions are chosen as the same as the above simulation setting, which can avoid the reflected formation. The trajectories of each agent and the final shape are depicted in Fig. \ref{fig:orientation_with_centroid}, which clearly show that the desired formation shape with the correct orientation is achieved and the formation centroid is preserved. The trajectories of each distance error and the orientation error for the edge $(1,2)$ are depicted in Fig. \ref{fig:orientation_with_centroid_errors}, which show an exponential convergence to the desired formation shape.



\begin{figure}
  \centering
  \includegraphics[width=60mm]{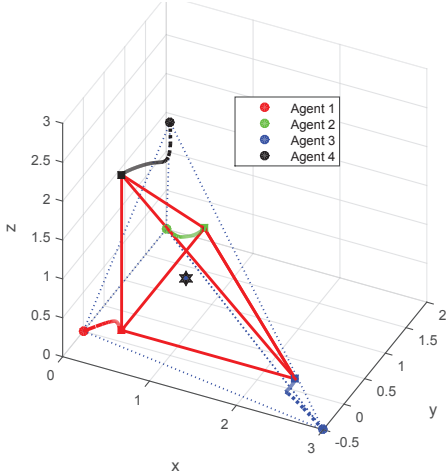}
  \caption{Stabilization of  a 3-D rigid formation with prescribed orientation. The initial and final positions are denoted by circles and squares, respectively. The initial formation is denoted by dotted blue lines, and the final formation is denoted by pink solid lines. The red line denotes the formation centroid. }
  \label{fig:3D}
\end{figure}


Lastly we show an example of stabilizing a rigid 3-D formation with desired orientation. The target formation is a tetrahedron, with  the desired distances given by $d_{12}^* =d_{13}^* =d_{14}^*= 2$, $d_{23}^* = d_{34}^* = d_{24}^*=2\sqrt{2}$. The desired orientation is that the edges  $(1,3)$  and $(1,4)$ should be aligned with the   the $x$-axis and the $z$-axis, respectively, which defines the orientation for the target tetrahedron formation.  Following the control strategy in Section 3, the desired relative position vectors for edges $(1,3)$  and $(1,4)$ are defined as $\hat p_1 - \hat p_2 = (2,0, 0)^T$ and $\hat p_1 - \hat p_4 = (0,0, 2)^T$, in which agents 1, 2 and 4 are chosen as orientation agents. The formation convergence is depicted in Fig. \ref{fig:3D}, which shows the successful achievement of the formation control task with both desired rigid shape and formation orientation.

\section{Conclusion}
In this paper we have discussed the formation control problem to achieve both   desired rigid shapes and   formation orientation. The designed controller combines the advantages of displacement-based approach and distance-based approach, by specifying  a small number of agents  as orientation agents which are tasked to control  relative position vectors associated with them to  desired directions. The proposed controllers are distributed in   that only relative measurements from neighboring agents are required.  For all non-orientation agents, any information about the global coordinate system is not required for them to implement the control, which guarantees a minimal requirement of the global knowledge of the global coordinate system. Certain simulation examples are provided to demonstrate the effectiveness of the proposed formation controllers.

\begin{ack}                               

The work of B. D. O. Anderson and Z. Sun was supported
by NICTA, which is funded by the Australian Government
through the ICT Centre of Excellence program, and by the
Australian Research Council under grant DP130103610
and DP160104500. Z. Sun is also supported by the Prime
Minister’s Australia Asia Incoming Endeavour Postgraduate
Award from Australian Government. The work of
M.-C. Park and H.-S. Ahn was supported by the National
Research Foundation of Korea (NRF) funded by
the Ministry of Education, Science and Technology (NRF-
2015M2A8A4049953).

\end{ack}

\section*{Appendix: proofs of several lemmas}

We first show a useful result on the dimension of the null space for two matrices and their   product.
\begin{lemma} \label{lemma_nullAB}
Consider two matrices $A \in \mathbb{R}^{m\times n}$ and $B \in \mathbb{R}^{n\times k}$ and the matrix product $C := AB$. Then there holds $\text{dim}(\text{null}(C)) = \text{dim}(\text{null}(B)) +  \text{dim}(  (\text{null}(A)  \cap \text{Im}(B)) $.
\end{lemma}
\textbf{Proof}
According to  the Sylvester Rank Theorem \cite[Theorem 2.6]{zhang2011matrix}, there holds $\text{rank}(C) = \text{rank}(B) - \text{dim}((\text{null}(A)  \cap \text{Im}(B))$. Also, from the fundamental rank-nullity theorem one obtains $\text{rank}(C) = k - \text{dim}(\text{null}(C))$ and $\text{rank}(B) = k - \text{dim}(\text{null}(B))$, which implies the desired result.
\qed

The following corollary is a consequence of Lemma \ref{lemma_nullAB}.
\begin{corollary}\label{corollary_nullAB}
Consider two matrices $A \in \mathbb{R}^{m\times n}$ and $B \in \mathbb{R}^{n\times k}$ and the matrix product $C := AB$. If $\text{dim}(  (\text{null}(A)  \cap \text{Im}(B)) = 0$, then $\text{null}(C) = \text{null}(B)$.
\end{corollary}
Lemma \ref{lemma_nullAB} and Corollary  \ref{corollary_nullAB} will be used later for analyzing the structure of null space of a rigidity matrix. 

\textbf{Proof of Lemma \ref{lemma:null_vector}}
The results can be verified by direct calculations.  Here we aim to provide a full proof to show how to construct these null vectors, as the analysis will be useful in later proofs for some key lemmas.

From the assumption that the target formation is infinitesimally rigid, one has $\text{rank}(\text{null}(R^TR)) = \text{rank}(\text{null}(R)) = {d(d+1)}/{2}$. The ${d(d+1)}/{2}$ dimensional null space contains  infinitesimal motions which preserve  inter-agent distances, with dimension $d$ corresponding to the translational motion, and  dimension ${d(d-1)}/{2}$ corresponding to the rotational motion \cite{hendrickson1992conditions}. Also note that $R = Z^T (H \otimes I_d)$. Since the target formation is assumed to be infinitesimally rigid, the underlying graph for shape control should be at least connected, which implies that $\text{null}(H \otimes I_d) = \text{span} ({\bf {1}}_n \otimes I_d )$. Then according to Lemma \ref{lemma_nullAB}, there holds $\text{null}(H \otimes I_d) \subset \text{null}(R) $, which implies that  $\text{span}({\bf 1}_n \otimes I_d)$ is a $d$-dimensional subspace of the null space of $R$ corresponding to the translational motion. This proves that the null vectors $q_1, q_2$ are valid bases in the null space of $R$ for the 2-D formation case and $q_1, q_2, q_3$ are valid bases in the null space of $R$ for the 3-D formation case.

We now divide the rest of the proof in the following two parts, according to the space dimensions:
\begin{itemize}
\item  The case of $d=2$:
\\
It is obvious  that $$Q_3 : = [(K_3 z_1)^T, (K_3 z_2)^T, \cdots, (K_3 z_m)^T]^T$$ is a null vector of $Z^T$, i.e. $\text{span}(Q_3) \subset \text{null} (Z^T) = \text{null} (ZZ^T)$. Also note that
\begin{align} \label{eq:transformation_image_null}
Q_3 & = (I_m \otimes K_3)z = (I_m \otimes K_3) (H \otimes I_d) p  \nonumber \\
&= (H \otimes K_3) p = (H \otimes I_d) (I_n \otimes K_3) p   
\end{align}
which means that $\text{span}(Q_3)$ is in the image of $(H \otimes I_d)$. Thus according to Lemma \ref{lemma_nullAB}, $(I_n \otimes K_3)   p : = q_3$ is a null vector of the rigidity matrix $R$. Note that the null space of $R$ corresponding to the rotational invariance is of dimension 1, which implies that $q_3$ is \emph{the unique} vector basis corresponding to the infinitesimal rotational motion.

\item  The case of $d=3$:
\\
It is obvious that the three vectors \begin{align}
Q_4 : = &[(K_4 z_1)^T, (K_4 z_2)^T, \cdots, (K_4 z_m)^T]^T  \nonumber \\
= & (I_m \otimes K_4)z =  (I_m \otimes K_4) (H \otimes I_d) p \nonumber \\
Q_5 : = & [(K_5 z_1)^T, (K_5 z_2)^T, \cdots, (K_5 z_m)^T]^T  \nonumber \\
= & (I_m \otimes K_5)z  =  (I_m \otimes K_5) (H \otimes I_d) p \nonumber \\
Q_6 : = & [(K_6 z_1)^T, (K_6 z_2)^T, \cdots, (K_6 z_m)^T]^T   \nonumber \\
= & (I_m \otimes K_6)z =  (I_m \otimes K_6) (H \otimes I_d) p \nonumber  
\end{align}
 are linearly independent, and are three bases of null spaces of $Z^T$. Similarly to Eq. \eqref{eq:transformation_image_null}, $Q_j$ can be rewritten as $$Q_j   = (H \otimes I_d) (I_n \otimes K_j) p$$ for $j = 4,5,6$, which implies that $\text{span}(Q_4, Q_5, Q_6)$ is in the image of $(H \otimes I_d)$.  Note that the null space of $R$ corresponding to the rotational invariance is of dimension $\frac{3\times2}{2} = 3$.  Thus according to Lemma \ref{lemma_nullAB}, the three vectors $(I_n \otimes K_j) p : = q_j$ for $j = 4,5,6$ are \emph{the} three null vector bases of $R$ corresponding to the infinitesimal rotational motion.
\end{itemize}
The proof is completed. \qed

\textbf{Proof of Lemma \ref{lemma:noncollinear_noncomplanar}}
The proof is done by trying to construct some null vectors of $R$ which are not in the null space spanned by the derived null vectors shown in Lemma \ref{lemma:null_vector}, thus leading to contradictions to the rank condition of the infinitesimal rigidity in Theorem 1.
\begin{itemize}
\item The case of $d=2$:
\\
The infinitesimal rigidity excludes the case that  $p_j - p_i = \mathbf{0}$ for all $j \in \mathcal{N}_i$. Now we  suppose that at least one $p_j - p_i$ is non-zero, and all the other $p_k - p_i$, $k \in \mathcal{N}_i$ can be described by a linear weight  of $p_j - p_i$. Then construct a vector $\bar q = [\mathbf{0}^T, \mathbf{0}^T, \cdots, (K_3(p_j - p_i))^T, \cdots, \mathbf{0}^T]^T$ with the non-zero term $K_3(p_j - p_i)$ in the $2(i-1)+1$ to $2i$ block. It is obvious  that $R\bar q=0$ and $\bar q \not\in \text{span}(q_1, q_2, q_3)$. Thus, the existence of the null vector $\bar q$ in this case increases the dimension of the null space of $R$ and therefore $\text{rank}(R) < 2n-3$, which violates the assumption that the framework is  infinitesimally rigid. In conclusion, the set of relative position vectors $p_j - p_i$, $j \in \mathcal{N}_i$ cannot all be collinear for any $i$.

\item The case of $d=3$:
\\
Similarly to the $d=2$ case, the infinitesimal rigidity excludes the case that all $p_j - p_i = \mathbf{0}$ for all $j \in \mathcal{N}_i$. The case that all $p_j - p_i$, $j \in \mathcal{N}_i$ are linearly dependent can be excluded by using the same argument as above. Now we suppose that   two $p_j - p_i, p_{j'} - p_i$, $j, j' \in \mathcal{N}_i$ are non-zero, and all the other $p_k - p_i$, $k \in \mathcal{N}_i$ can be described as linear combinations of these two. Then construct a vector $\bar q = [\mathbf{0}^T, \mathbf{0}^T, \cdots, ((p_j - p_i)\wedge(p_{j'} - p_i))^T, \cdots, \mathbf{0}^T]^T$, with the non-zero term $(p_j - p_i)\wedge(p_{j'} - p_i)$ in the $3(i-1)+1$ to $3i$ block. By direct calculations, it can be shown  that $R\bar q=0$ and $\bar q \not\in \text{span}(q_1, q_2, q_3, q_4, q_5, q_6)$. Thus, the existence of the null vector $\bar q$ in this case increases the dimension of the null space of $R$ and therefore $\text{rank}(R) < 3n-6$, which violates the assumption that the framework is  infinitesimally rigid. In conclusion, the set of relative position vectors $p_j - p_i$, $j \in \mathcal{N}_i$ cannot all be coplanar for any $i$.
\end{itemize}
By summarizing the above arguments, the proof is completed.
\qed

\textbf{Proof of Lemma \ref{lemma:null_vector_F}}
First note that both $R^TR$ and $L_o \otimes I_d$ are symmetric and positive semidefinite.   From Lemma \ref{lemma:null_vector} one knows that $\text{span}({\bf 1}_n \otimes I_d)$ is a subspace of the null space of $R$. Also, $\text{span}({\bf 1}_n \otimes I_d)$ is a subspace of the null space of $L_o \otimes I_d$. Thus, there holds $\text{span}({\bf 1}_n \otimes I_d) \subset \text{null}(\mathcal{F})$.
We then show that there does not exist other null vectors in $\text{null}(\mathcal{F})$.



We introduce a selection matrix, denoted by $J \in \mathbb{R}^{m \times m}$, whose $k$-th row is $e_k$ (i.e. the $k$-th standard basis) if the $k$-th edge is selected as the orientation edge, or the $k$-th row is an all-zero vector otherwise. Note that $J^T = J$. Denote the incidence matrix for the underlying graph of orientation control as $H_o$. By doing this, there holds $H_o = J H$ and thus $L_o = H_o^T H_o = H^T  J  J H$ since the  underlying graph of orientation control is assumed to be undirected. Thus $\mathcal{F} = (H \otimes I_d)^T  Z Z^T (H \otimes I_d) + (H \otimes I_d)^T  (J J \otimes I_d) (H \otimes I_d) = (H \otimes I_d)^T  (Z Z^T + J  J \otimes I_d ) (H \otimes I_d)$. We now divide the proof in the following two parts, according to the space dimensions:

\begin{itemize}
\item  The case of $d=2$:
\\
From Lemma \ref{lemma:null_vector},  $Q_3$  is a null vector of $Z^T$ and $q_3$ is   a null vector of $R$.
By direct calculation, it holds $(Z Z^T + J  J \otimes I_2 )Q_3 \neq 0$, i.e. $Q_3$ is not a null vector of the matrix $(Z Z^T + J  J \otimes I_2)$, which together with Corollary \ref{corollary_nullAB} implies that $q_3$ is not a null vector to the matrix $\mathcal{F}$.  Thus, there holds $\text{null}(\mathcal{F})  = \text{span} ({\bf {1}} \otimes I_2 )$, which implies that the null space of  $\mathcal{F}$ is of dimension 2 and $\mathcal{F}$ has 2 zero eigenvalues.

\item  The case of $d=3$:
\\
We fix $j = 4,5,6$. In this case, there are at least two non-zero rows in $J$, corresponding to at least two adjacent edges selected in the orientation control. From Lemma \ref{lemma:null_vector}, each $Q_j$ is a null vector of $Z^T$ and each $q_j$ is also a null vector  of $R$. Following similar steps as above for the 2-D case and by direct calculation, it holds that   $(Z Z^T + J  J \otimes I_3 )Q_j \neq 0$. Thus $Q_j$ are not  null vectors of $(Z Z^T + J  J \otimes I_3 )$, which together with Corollary \ref{corollary_nullAB} implies $q_j$ are not null vectors of the matrix $\mathcal{F}$.  Thus, there holds $\text{null}(\mathcal{F})   = \text{span} ({\bf {1}} \otimes I_3 )$, which implies that the null space of  $\mathcal{F}$ is of dimension 3 and $\mathcal{F}$ has 3 zero eigenvalues.
\end{itemize}
By summarizing the above arguments, the statements in the lemma are proved.
\qed

    \bibliographystyle{elsarticle-harv}
\bibliography{Formation_attitude_articles}        

\begin{thebibliography}{33}
\expandafter\ifx\csname natexlab\endcsname\relax\def\natexlab#1{#1}\fi
\expandafter\ifx\csname url\endcsname\relax
  \def\url#1{\texttt{#1}}\fi
\expandafter\ifx\csname urlprefix\endcsname\relax\def\urlprefix{URL }\fi

\bibitem[{Anderson et~al.(2010)Anderson, Shames, Mao, and
  Fidan}]{anderson2010formal}
Anderson, B.~D., Shames, I., Mao, G., Fidan, B., 2010. Formal theory of noisy
  sensor network localization. SIAM Journal on Discrete Mathematics 24~(2),
  684--698.

\bibitem[{Anderson and Helmke(2014)}]{anderson2014counting}
Anderson, B. D.~O., Helmke, U., 2014. Counting critical formations on a line.
  SIAM Journal on Control and Optimization 52~(1), 219--242.

\bibitem[{Aranda et~al.(2015)Aranda, L{\'o}pez-Nicol{\'a}s, Sag{\"u}{\'e}s, and
  Zavlanos}]{aranda2015coordinate}
Aranda, M., L{\'o}pez-Nicol{\'a}s, G., Sag{\"u}{\'e}s, C., Zavlanos, M.~M.,
  2015. Coordinate-free formation stabilization based on relative position
  measurements. Automatica 57, 11--20.

\bibitem[{Asimow and Roth(1979)}]{asimow1979rigidity}
Asimow, L., Roth, B., 1979. {The rigidity of graphs, II}. Journal of
  Mathematical Analysis and Applications 68~(1), 171--190.

\bibitem[{Cai and De~Queiroz(2015)}]{cai2015adaptive}
Cai, X., De~Queiroz, M., 2015. Adaptive rigidity-based formation control for
  multirobotic vehicles with dynamics. IEEE Transactions on Control Systems
  Technology 23~(1), 389--396.

\bibitem[{Cao et~al.(2011)Cao, Morse, Yu, Anderson, and
  Dasgupta}]{cao2011maintaining}
Cao, M., Morse, A.~S., Yu, C., Anderson, B. D.~O., Dasgupta, S., 2011.
  Maintaining a directed, triangular formation of mobile autonomous agents.
  Communications in Information and Systems 11~(1), 1--16.

\bibitem[{Cort{\'e}s(2009)}]{cortes2009global}
Cort{\'e}s, J., 2009. Global and robust formation-shape stabilization of
  relative sensing networks. Automatica 45~(12), 2754--2762.

\bibitem[{Dorfler and Francis(2010)}]{dorfler2010geometric}
Dorfler, F., Francis, B., 2010. Geometric analysis of the formation problem for
  autonomous robots. IEEE Transactions on Automatic Control 55~(10),
  2379--2384.

\bibitem[{{Garcia de Marina} et~al.(2016){Garcia de Marina}, {Jayawardhana},
  and {Cao}}]{Hector2016maneuvering}
{Garcia de Marina}, H., {Jayawardhana}, B., {Cao}, M., 2016. Distributed
  rotational and translational maneuvering of rigid formations and its
  applications. IEEE Transactions on Robotics. Accepted and in press. DOI:
  10.1109/TRO.2016.2559511. Also vailabe at arXiv:1604.07849.

\bibitem[{Hendrickson(1992)}]{hendrickson1992conditions}
Hendrickson, B., 1992. Conditions for unique graph realizations. SIAM Journal
  on Computing 21~(1), 65--84.

\bibitem[{Khalil(1996)}]{khalil1996nonlinear}
Khalil, H.~K., 1996. Nonlinear systems. Vol.~3. Prentice hall New Jersey.

\bibitem[{Krick et~al.(2009)Krick, Broucke, and
  Francis}]{krick2009stabilisation}
Krick, L., Broucke, M.~E., Francis, B.~A., 2009. Stabilisation of
  infinitesimally rigid formations of multi-robot networks. International
  Journal of Control 82~(3), 423--439.

\bibitem[{Lin et~al.(2015)Lin, Wang, Chen, Fu, and Han}]{Lin_affine_formation}
Lin, Z., Wang, L., Chen, Z., Fu, M., Han, Z., 2015. Necessary and sufficient
  graphical conditions for affine formation control. IEEE Transactions on
  Automatic Control, in press, 10.1109/TAC.2015.2504265.

\bibitem[{Markdahl et~al.(2012)Markdahl, Karayiannidis, Hu, and
  Kragic}]{markdahl2012distributed}
Markdahl, J., Karayiannidis, Y., Hu, X., Kragic, D., 2012. Distributed
  cooperative object attitude manipulation. In: Proc. of IEEE International
  Conference on Robotics and Automation (ICRA). IEEE, pp. 2960--2965.

\bibitem[{Meng et~al.(2016)Meng, Anderson, and Hirche}]{Meng2014Compass}
Meng, Z., Anderson, B. D.~O., Hirche, S., 2016. Formation control with
  mismatched compasses. Automatica 69, 232--241.

\bibitem[{Montijano et~al.(2014)Montijano, Zhou, Schwager, and
  Sagues}]{Montijano2014}
Montijano, E., Zhou, D., Schwager, M., Sagues, C., June 2014. Distributed
  formation control without a global reference frame. In: Proc. of the 2014
  American Control Conference. pp. 3862--3867.

\bibitem[{Mou et~al.(2016)Mou, Morse, Belabbas, Sun, and
  Anderson}]{SMA16TACsub}
Mou, S., Morse, A.~S., Belabbas, M.~A., Sun, Z., Anderson, B. D.~O., 2016.
  Undirected rigid formations are problematic. IEEE Transactions on Automatic
  Control. Accepted and in press. DOI: 10.1109/TAC.2015.2504479.

\bibitem[{Oh and Ahn(2011)}]{oh2011formation}
Oh, K.-K., Ahn, H.-S., 2011. Formation control of mobile agents based on
  inter-agent distance dynamics. Automatica 47~(10), 2306--2312.

\bibitem[{Oh and Ahn(2014{\natexlab{a}})}]{oh2014distance}
Oh, K.-K., Ahn, H.-S., 2014{\natexlab{a}}. Distance-based undirected formations
  of single-integrator and double-integrator modeled agents in n-dimensional
  space. International Journal of Robust and Nonlinear Control 24~(12),
  1809--1820.

\bibitem[{Oh and Ahn(2014{\natexlab{b}})}]{Oh2014Orientation}
Oh, K.-K., Ahn, H.-S., Feb 2014{\natexlab{b}}. Formation control and network
  localization via orientation alignment. IEEE Transactions on Automatic
  Control 59~(2), 540--545.

\bibitem[{Oh et~al.(2015)Oh, Park, and Ahn}]{oh2014survey}
Oh, K.-K., Park, M.-C., Ahn, H.-S., 2015. A survey of multi-agent formation
  control. Automatica 53, 424 -- 440.

\bibitem[{Pais et~al.(2009)Pais, Cao, and Leonard}]{pais2009formation}
Pais, D., Cao, M., Leonard, N.~E., 2009. Formation shape and orientation
  control using projected collinear tensegrity structures. In: Proc. of the
  2009 American Control Conference. IEEE, pp. 610--615.

\bibitem[{Park and Ahn(2015)}]{myoung2015CDC}
Park, M.-C., Ahn, H.-S., 2015. Distance-based control of formations with
  orientation control. In: Proc. of the 54th IEEE Conference on Decision and
  Control. pp. 2199--2104.

\bibitem[{Ren and Beard(2008)}]{ren2008distributed}
Ren, W., Beard, R.~W., 2008. Distributed consensus in multi-vehicle cooperative
  control. Springer.

\bibitem[{Sun and Anderson(2015)}]{sun2015orientation}
Sun, Z., Anderson, B. D.~O., 2015. Rigid formation control with prescribed
  orientation. In: Proc. of the IEEE Multi-Conference on Systems and Control
  (MSC’15). pp. 639--645.

\bibitem[{Sun et~al.(2014)Sun, Mou, Anderson, and Morse}]{Sun2014CDC}
Sun, Z., Mou, S., Anderson, B. D.~O., Morse, A.~S., 2014. Formation movements
  in minimally rigid formation control with mismatched mutual distances. In:
  Proc. of the 53rd Conference on Decision and Control. IEEE, pp. 6161--6166.

\bibitem[{Tay and Whiteley(1985)}]{tay1985generating}
Tay, T.-S., Whiteley, W., 1985. Generating isostatic frameworks. Structural
  Topology 1985 11, 21--49.

\bibitem[{Tian and Wang(2013)}]{tian2013global}
Tian, Y.-P., Wang, Q., 2013. Global stabilization of rigid formations in the
  plane. Automatica 49~(5), 1436--1441.

\bibitem[{Wang et~al.(2011)Wang, Markdahl, and Hu}]{wang2011distributed}
Wang, L., Markdahl, J., Hu, X., 2011. Distributed attitude control of
  multi-agent formations. In: Proc. of the 18th IFAC World Congress. pp.
  2965--2971.

\bibitem[{Xiao et~al.(2009)Xiao, Wang, Chen, and Gao}]{xiao2009finite}
Xiao, F., Wang, L., Chen, J., Gao, Y., 2009. Finite-time formation control for
  multi-agent systems. Automatica 45~(11), 2605--2611.

\bibitem[{Zelazo et~al.(2015)Zelazo, Franchi, Bülthoff, and
  Robuffo~Giordano}]{Zelazo01012015}
Zelazo, D., Franchi, A., Bülthoff, H.~H., Robuffo~Giordano, P., 2015.
  Decentralized rigidity maintenance control with range measurements for
  multi-robot systems. The International Journal of Robotics Research 34~(1),
  105--128.

\bibitem[{Zhang(2011)}]{zhang2011matrix}
Zhang, F., 2011. Matrix theory: basic results and techniques. Springer Science
  \& Business Media.

\bibitem[{Zhao and Zelazo(2016)}]{zhao2014bearing}
Zhao, S., Zelazo, D., 2016. Bearing rigidity and almost global bearing-only
  formation stabilization. IEEE Transactions on Automatic Control 61~(5),
  1255--1268.

\end{thebibliography}

\end{document}